\documentclass[fleqn,usenatbib]{mnras}

\usepackage{newtxtext,newtxmath}
\usepackage[T1]{fontenc}

\DeclareRobustCommand{\VAN}[3]{#2}
\let\VANthebibliography\thebibliography
\def\thebibliography{\DeclareRobustCommand{\VAN}[3]{##3}\VANthebibliography}

\usepackage{graphicx}	% Including figure files
\usepackage{amsmath}	% Advanced maths commands
\usepackage{makecell}
\usepackage[flushleft]{threeparttable}
\title[QPOs in 1RXS~J180408.9$-$342058]{Detection of Millihertz Quasi-Periodic Oscillations in the X-Ray Binary 1RXS~J180408.9$-$342058
}

\author[K. Tse et al.]{Kaho Tse$^{1-3}$\thanks{E-mail: kaho.tse@monash.edu},
Duncan K. Galloway$^{1,4,5}$,
Yi Chou$^{2}$,
Alexander Heger$^{1,4-6}$,
and \newauthor Hung-En Hsieh$^{2}$
\\
$^1$School of Physics and Astronomy, Monash University, Victoria 3800, Australia\\
$^2$Graduate Institute of Astronomy, National Central University, Jhongli 32001, Taiwan\\
$^3$Department of Physics, National Central University, Jhongli 32001, Taiwan\\
$^4$OzGrav-Monash, School of Physics and Astronomy, Monash University, VIC 3800, Australia\\
$^5$Joint Institute for Nuclear Astrophysics - Center for the Evolution of the Elements (JINA-CEE), Monash University, Vic 3800, Australia\\
$^6$ARC Center of Excellence for Astrophysics in Three Dimensions (ASTRO-3D), Australia\\
}

\date{Accepted XXX. Received YYY; in original form ZZZ}

\pubyear{2020}

\begin{document}
\label{firstpage}
\pagerange{\pageref{firstpage}--\pageref{lastpage}}
\maketitle
\begin{abstract}
% general suggestion - avoid acronyms in abstract -- dkg
Millihertz quasi-periodic oscillations (mHz QPOs) observed in neutron-star low-mass X-ray binaries (NS LMXBs) are generally explained as marginally stable thermonuclear burning on the neutron star surface.  We report the discovery of mHz QPOs in an \textsl{XMM-Newton} observation of the transient 1RXS~J180408.9$-$342058, during a regular bursting phase of its 2015 outburst.  We found significant periodic signals in the March observation, with frequencies in the range 5--8$\,\mathrm{mHz}$, superimposed on a strong $\sim1/f$ power-law noise continuum.  Neither the QPO signals nor the power-law noise were present during the April observation, which exhibited a $2.5\times$ higher luminosity and had correspondingly more frequent bursts.  When present, the QPO signal power decreases during bursts and disappears afterwards, similar to the behaviour in other sources.  1RXS~J180408.9$-$342058 is the eighth source known to date that exhibits such QPOs driven by thermonuclear burning.  We examine the range of properties of the QPO signals in different sources.  Whereas the observed oscillation profile is similar to that predicted by numerical models, the amplitudes are significantly higher, challenging their explanation as originating from marginally stable burning.
\end{abstract}
 

%%%%%%%%%%%%%%%%%%%%%%%%%%%%%%%%%%%%%%%%%%%%%%%%%%

%%%%%%%%%%%%%%%%% BODY OF PAPER %%%%%%%%%%%%%%%%%%

\section{Introduction}
Type I X-ray bursts are thermonuclear-powered flashes on the surface of neutron stars in low mass X-ray binaries.  These energetic outbursts are triggered by a thermonuclear runaway process.  Through Rocbe-Lobe overflow, material accretes from the donor onto the neutron star and is then compressed and heated under strong ($\sim10^{14}\ \rm{g\,cm^{-2}}$) gravity.  Upon reaching the ignition condition, thermonuclear runaway is triggered, usually by the triple$-\alpha$ reaction, and the $\alpha p$  process operates to burn helium up to the iron group.  Followed by the \textsl{rp} process (rapid proton capture process), hydrogen is burnt close to the proton drip line to produce heavier elements and release additional energy.  The bursts typically reach about $10^{38}\,\mathrm{erg}\,\mathrm{s}^{-1}$, and last for some ten to a hundred seconds, depending mainly on their hydrogen and helium mass fractions (see \citealt{2003astro.ph..1544S} and \citealt{2017arXiv171206227G} for reviews).

\cite{firstqpo} first discovered $\approx8$ mHz quasi-periodic oscillations in the sources 4U~1636$-$536, 4U~1608$-$52 as well as 
% note that the hyphen in names like Aql X-1 is a hyphen not a minus sign!
% (just to be extra confusing)
Aql~X-1, and suggested that the signals may originate from a mode of nuclear burning on the neutron star surface.  \cite{heger2007} later found a particular burning regime showing an oscillatory behaviour at an accretion rate of $0.925\,\dot{M}_{\text{Edd}}$, where $\dot{M}_{\text{Edd}}=1.75\times10^{-8}\times1.7/(X+1)\,\text{M}_{\odot}\text{yr}^{-1}$ is the Eddington-limited rate with the hydrogen mass fraction $X$, in their simulations using the 1D stellar hydrodynamics code \textsc{Kepler} \citep{1978ApJ...225.1021W,2004ApJS..151...75W}.  They interpreted this oscillation, which takes place within a narrow range of accretion rates, as the transitional phase between stable and unstable nuclear burning, as also predicted by one zone models (e.g., \citealt{1983ApJ...264..282P}).  In this regime,
%being close to the critical accretion rate, 
the thermonuclear generation rate is comparable to the cooling rate, which leads to this oscillatory behaviour. The oscillation period, $P\simeq\sqrt{t_\text{therm}\,t_\text{acc}}$, where $t_\text{therm}$ and $t_\text{acc}$ represent the time for gas to cool (thermal) and for replenishing fuel (accretion) respectively.  This period found in the \cite{heger2007} simulations is consistent with the reported range of frequencies of $\mathrm{mHz}$ QPOs. 

\begin{figure*}
	\includegraphics[width=.7\textwidth]{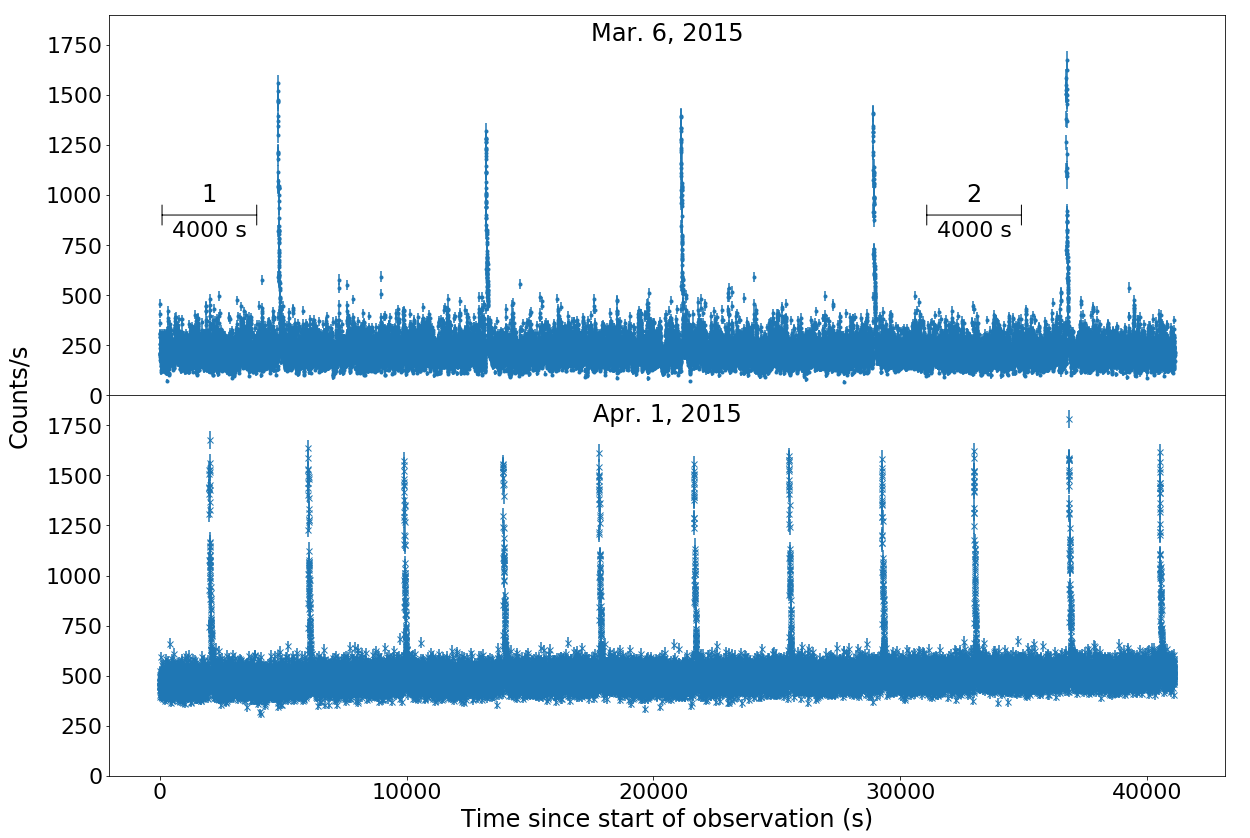}
    \caption{The $0.7$--$10\,\mathrm{keV}$, $1\,\mathrm{s}$ bin size light curves of 1RXS~J180408.9$-$342058 from the two \textsl{XMM-Newton} observations on 2015 March 6 and 2015 April 1, respectively.  The thermonuclear bursts are evident as the brief increases of the count rate above the persistent level.  The burst rate and persistent count rate increase by about two times in the second observation period (\textsl{bottom panel}).  Significant periodic signals appear in the labeled segments from the first observation, whereas no periodicity was detected in the second observation.}
    \label{fig:lc}
\end{figure*}

mHz QPOs have been detected in several other sources \cite[]{2011ATel.3258....1S,2012ApJ...748...82L,2018ApJ...865...63S,2019MNRAS.486L..74M}, that are evidently related to the thermonuclear burning on the neutron star surface, as Type I X-ray bursts are also present in the observations.  Although there are discrepancies between the observations and theory, for instance the disagreement in the accretion rate for the oscillations to exist, additional samples could enhance our understanding of the marginally stable burning or even put constraints on NS parameters, such as its equation of state \citep{2016ApJ...831...34S}.

1RXS~J180408.9$-$342058 ($l=357.30^{\circ}\,,b=-6.13^{\circ}$) was identified as a transient X-ray binary in 2012 when a Type I X-ray burst from the source was detected by \textsl{INTEGRAL}.  It is classified as an "atoll" source, which traces out that pattern in its colour-colour diagram \citep{atoll}.  The source is located at a distance thought to be at most $5.8\,\mathrm{kpc}$ by assuming the Eddington luminosity limit $=3.8\times10^{38}\,\mathrm{erg}\,\mathrm{s}^{-1}$ for a helium-rich burst \citep{2012ATel.4050....1C}.  \cite{2016A&A...587A.102B} initially suggested that this source is an ultra compact X-ray binary with a helium white dwarf companion, based on their estimation of the orbital period of this binary system, and the optical spectrum.  The hydrogen rich bursts observed by \textsl{NuSTAR} in 2015 \citep{2019MNRAS.490.2300M}, however, suggest that the source accretes mixed H/He fuel rather than pure helium.  The nature of this binary system is therefore needed to be further justified.  The atoll source was in a hard spectral state during an \textsl{XMM-Newton} observation in March \citep{2016ApJ...831...34S}, and showed similar variability characteristics to the ``extreme island'' state in its power density spectrum \citep{2017MNRAS.472..559W}.  Nearly a month later, the source, with slightly higher accretion rate, was identified being in the island state during the second \textsl{XMM-Newton} observation. 

In this paper, we report  timing analysis of \textsl{XMM-Newton} data  from 1RXS~J180408.9$-$342058.  We report the detection of significant periodic signals with frequency in the range $\sim5$ to $8\,\mathrm{mHz}$ from the source, that we attribute to the marginally stable thermonuclear burning.  We computed dynamical power spectra to show the drift of the oscillation frequencies, and explored how oscillation amplitude depends on energy. We compared the QPO profile from this work with a simulation result from \cite{heger2007}.  Finally, we discuss the possible reasons for the discrepancies between the observed mHz QPOs properties with those of numerical models. 

\section{observations and analysis}
The transient X-ray binary 1RXS~J180408.9$-$342058 was found in outburst by Swift/BAT on 2015 January 22 \citep[MJD 57045;][]{2015ATel.6997....1K} and remained active until 2015 April 3 \citep{2015ATel.7352....1D}.  \textit{XMM-Newton}\/ made two follow-up observations during this period.  The observations on 2015 March 6 (Obs.\ ID: 0741620101) and 2015 April 1 (Obs.\ ID: 0741620201) each have $57,\!000\,\mathrm{s}$ on-time, from which we report our analysis in this paper.  We used data taken from the timing mode of EPIC/pn \citep{2001A&A...365L..18S} and added the barycenter correction for all events by applying the task \texttt{barycorr} with the Science Analysis System (SAS) version \texttt{18.0.0}.  We followed the standard filtering of events for EPIC/pn, selecting single and double events.  We modelled a Gaussian function to the distribution of photons over the CCD and included the photons within $3\,\sigma$ as coming from the source, which accordingly covers the CCD columns $19 \leq\text{RAWX}\leq61$ and $27 \leq\text{RAWX}\leq52$ for the first and the second observations, respectively.  A $\text{RAWX}\leq15$ for the background applied for both observations.  We produced two background-subtracted light curves with $1\, \mathrm{s}$ bin size and $0.7$ to $10\,\mathrm{keV}$ energy range.

We used the Fourier Transform from the software package \textsc{stingray} \citep{2019ApJ...881...39H} to search for any periodic signals in every segment without bursts.  The power spectra were normalized following \cite{1983ApJ...266..160L} for which the Poisson noise level is expected to be equal to $2$ on average.

% PDS
In order to explore the possible frequency drift of the QPOs, we applied Lomb-Scargle periodograms \citep{1976Ap&SS..39..447L,1982ApJ...263..835S}, with $2,\!000\,\mathrm{s}$ sliding windows and $20\,\mathrm{s}$ for each forwarding step, to compute dynamical power spectra for all the burst-free segments.  We searched the range $5$ to $9\,\mathrm{mHz}$ and adopted the normalization from \cite{Press2007} for the power spectra.

% to result
% calculate limits

% pulse shape

% moved figure from end section to appear in line with the text

\begin{figure}
	\includegraphics[width=\columnwidth]{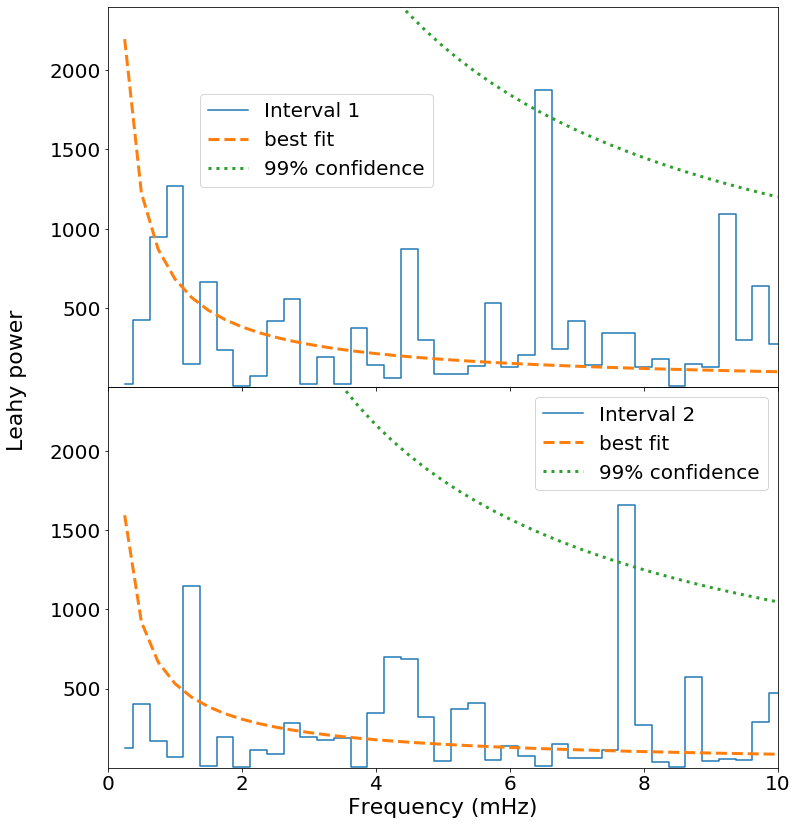}
    \caption{Two Fourier power spectra from the first and second $4,\!000\,\mathrm{s}$ intervals indicated with numbers in Figure~\ref{fig:lc}. The dashed and dotted lines show the fitted red noise continuum, and $99\,\%$ confidence level based on the null hypothesis under the red noise background respectively.  Strong peaks between $6$--$8\,\mathrm{mHz}$ appear on both spectra.}
    \label{fig:confidence}
\end{figure}

\section{results}
The resulting light curves from the both observations show regular Type I X-ray bursts (Figure~\ref{fig:lc}).  The burst rate in the second observation period is about twice as high as during the first period, and shows an enhanced persistent luminosity.  All bursts have a duration of $\sim150\,\mathrm{s}$, suggesting that they have a hydrogen-rich composition during burst.  We identified candidate signals in two intervals, as indicated in Figure~\ref{fig:lc}, labeled $1$ and $2$.  They show strong peaks in their power spectra at frequencies of $\approx6.5$ and $7.8\,\mathrm{mHz}$ respectively, but rather contaminated by red noise.  To determine the significance of these signals, we re-computed the power spectra for these two intervals, each with a $4,\!000\,\mathrm{s}$ continuous interval.  Following the procedure from \citet{2005A&A...431..391V} to test for the presence of periodicity with limited observational data against a red noise background, we first fit the two periodograms with a power-law model $P_i = Af^{-\alpha}_i$ with least squares regression to estimate the red noise continuum in the power spectra.  The frequency range of interest, $4$ to $10\,\mathrm{mHz}$, was ignored for the fit, such that the model is independent of the signals in this range of frequency.  The resulting exponents, $\alpha$, are $0.82\pm0.06$ and $0.78\pm0.06$ for the first and second intervals, respectively.  For a given modeled noise power $L_i$, we determined a $99\,\%$ confidence level based on the fact that the ratio of twice the signal power to the noise power, $2P_i/L_i$, is distributed as a $\chi^{2}_2$ nature over the periodograms as well as taking the number of trials into account.  Both of the peaks we detected exceed the $99\,\%$ confidence level (Figure~\ref{fig:confidence}).  We did not detect any periodic signals from the second observation where the $95\,\%$ confidence upper limit on the fractional RMS amplitude in the range $4$ to $10\,\mathrm{mHz}$ is $<0.62\,\%$.  

\begin{figure}
	\includegraphics[width=\columnwidth]{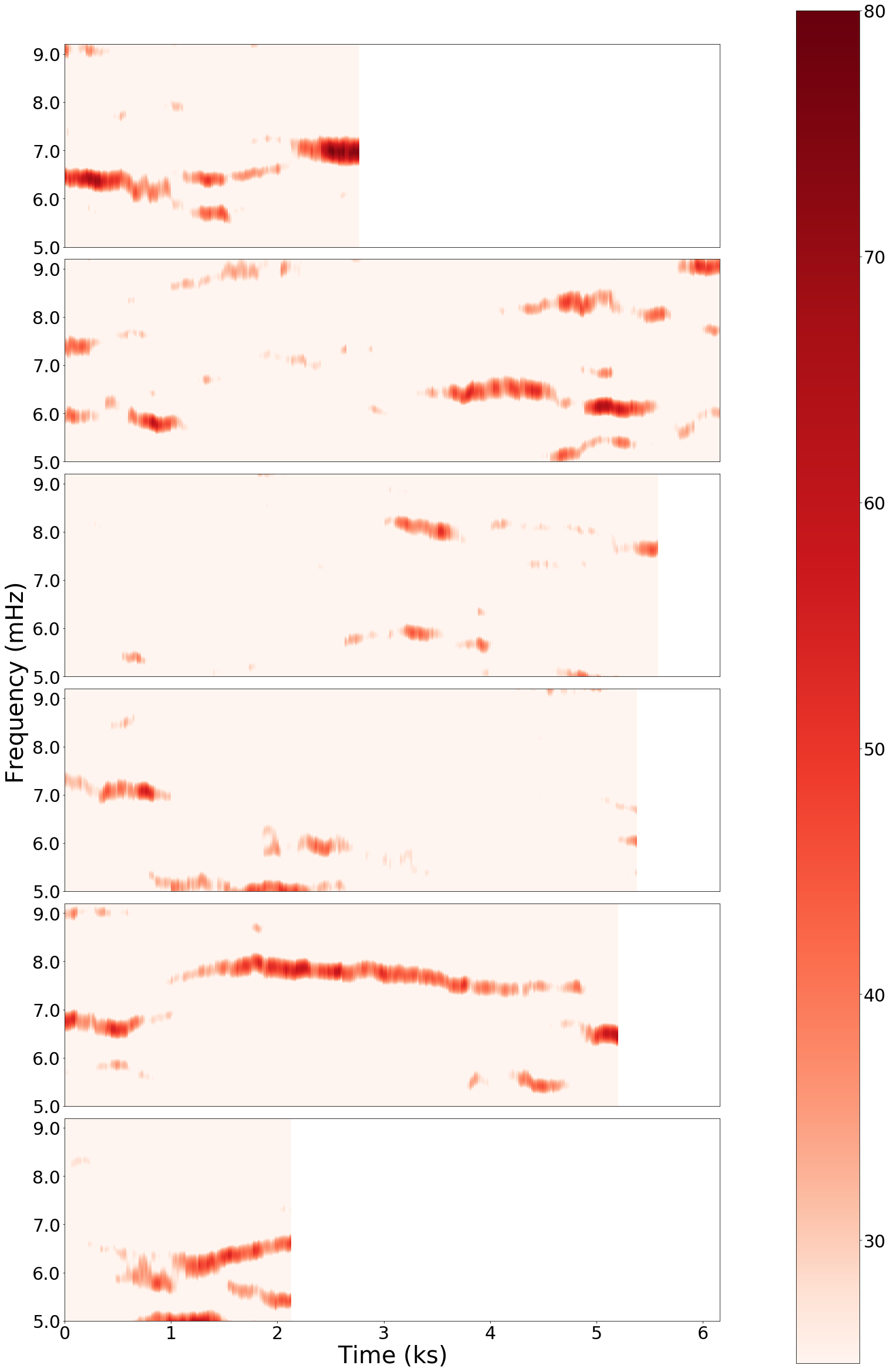}
    \caption{Dynamical power spectra showing the evolution of the $\mathrm{mHz}$ QPOs for all burst-free segments from the first observation using Lomb-Scargle periodograms with the normalization of \citep{Press2007},  $2,\!000\,\mathrm{s}$ sliding windows, and $20\,\mathrm{s}$ steps.  The end of each PDS is within a step of the onset of a burst.  For clarity, we plot only power $\geq25$ to filter out noise. The $x$-axis shows the start time of the current $2,\!000\,\mathrm{s}$ spectrum from the beginning of each segment.  The common color bar shown on the right hand side represents the signal power for all panels.} 
    \label{fig:pds}
\end{figure}

\begin{figure}
	\includegraphics[width=\columnwidth]{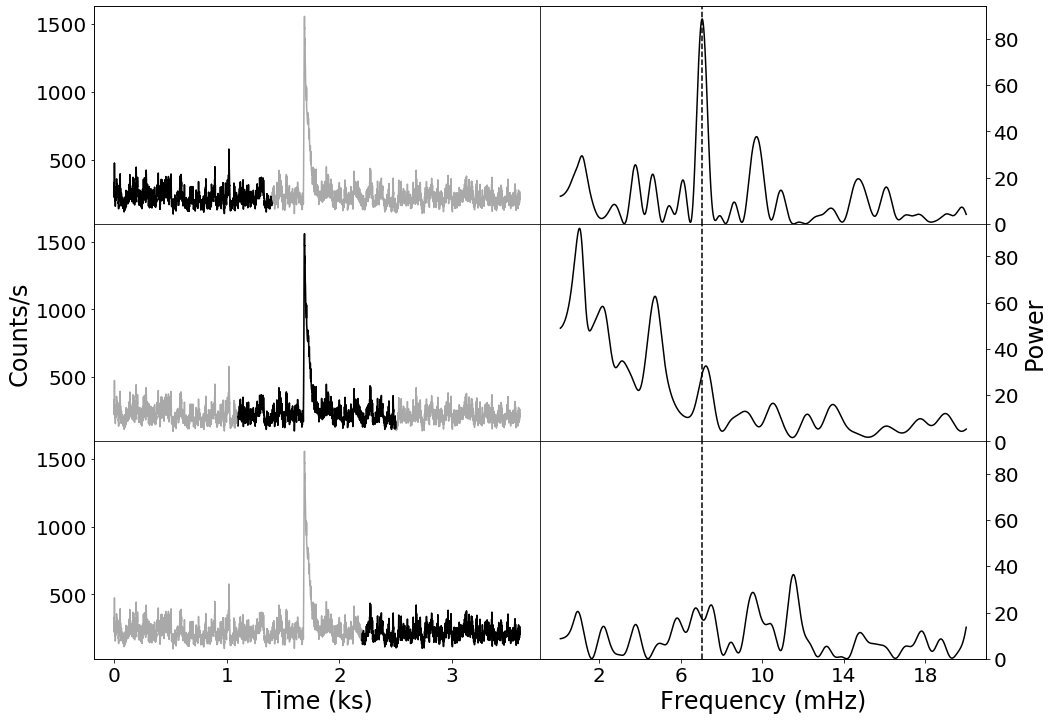}
    \caption{Three pairs of light curve and the corresponding power spectrum for selected time ranges.  The left panel, from top to bottom, shows three $1,\!400\,\mathrm{s}$ windows in dark color before, during and after the first observed Type I X-ray burst from the first observation, respectively.  A strong peak stands out at $\sim7\,\mathrm{mHz}$ before the flash.  Its power drops remarkably during the burst and being undetectable afterward.} 
    \label{fig:lc_with_spec}
\end{figure}

% new para here -- dkg
The dynamical power spectra of the six burst-free segments from the first observation show that the oscillations are time-dependent in both frequency and amplitude (Figure~\ref{fig:pds}).  Other panels also show high power signals in the periodgrams, but these are rather sporadic and only last for brief times.  To further investigated how the mHz QPOs correlate to the bursts, we have computed power spectra for light curve segments before, during, and after the flashes. Figure~\ref{fig:lc_with_spec} shows one example for the disappearance of mHz QPOs after the onset of the first Type I X-ray burst in the first observation.  Following the same normalization as above, %the Lomb-Scargle periodograms on the right hand side correspond to the $1,\!400\,\mathrm{s}$ windows, shown in dark color, of the light curves on the left hand side. 
the $\sim7\,\mathrm{mHz}$ periodic signal in the Lomb-Scargle periodograms, indicated by a vertically dashed line, drastically decreases in strength during the burst and becomes undetectable in the noise afterward, which implies the cessation of mHz QPOs.

\begin{figure}
	\includegraphics[width=\columnwidth]{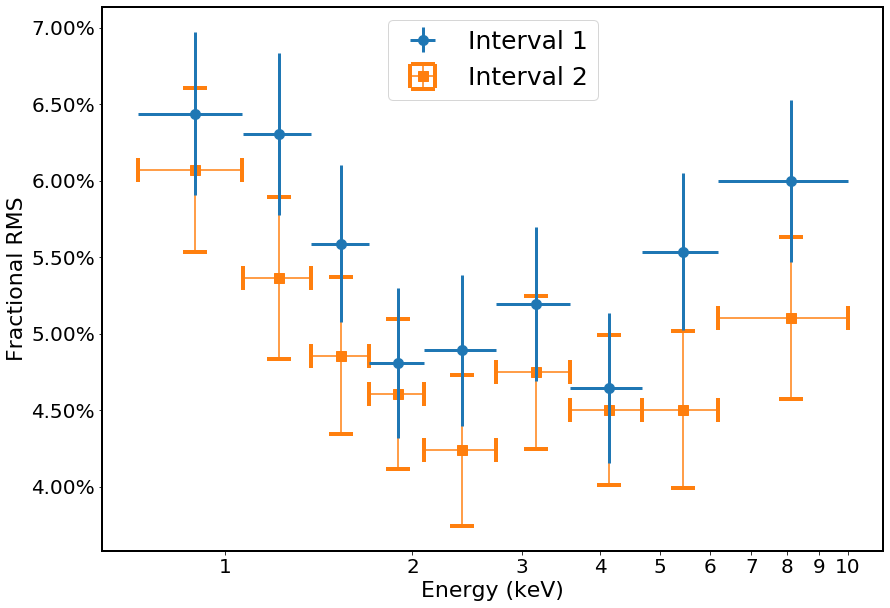}
    \caption{Fractional RMS amplitudes of $\mathrm{mHz}$ QPOs over energy bands from $0.7$ to $10\,\mathrm{keV}$.  We divided the events from the first (\textsl{blue circle}) and the second (\textsl{orange square}) $4,\!000\,\mathrm{s}$ continuous intervals in Figure~\ref{fig:lc} into 9 energy bands and folded them with their corresponding frequency, $6.57\,\mathrm{mHz}$ and $7.73\,\mathrm{mHz}$ respectively (see Figure~\ref{fig:confidence}).  Results from the two intervals show increases towards lower energy ($\lesssim 3\,\mathrm{keV}$).}
    \label{fig:energy_dependent}
\end{figure}

We used the above two $4,\!000\,\mathrm{s}$ segments labeled with numbers in Figure~\ref{fig:lc}, in which the detected oscillations have relatively stable frequencies and last for longer duration, to study the energy dependence on the oscillation amplitude.  The events from the two segments were divided into $9$ energy bands, from $0.7$ to $10\,\mathrm{keV}$, with roughly identical counts.  We folded the events in $16$ bins for each energy band, with their corresponding peak frequencies $6.57\,\mathrm{mHz}$ and $7.73\,\mathrm{mHz}$ from the result of the Fourier analysis respectively, to produce oscillation profiles and derived the fractional RMS amplitudes for the two segments.  The amplitude increases towards lower energy ($\lesssim 3\,\mathrm{keV}$) in both segments whereas the tendency of high energy part cannot be well resolved (Figure~\ref{fig:energy_dependent}).
We also compared the observed oscillation profile from the first segment in Figure~\ref{fig:lc} with a simulation result of mHz QPOs from \cite{heger2007}.  We recomputed the phase-folded QPOs profile of the combination of all energy ($0.7$ to $10\,\mathrm{keV}$) in $16$ bins for the segment and plotted referenced to the left $y$-axis (Figure~\ref{fig:overplot}).  For comparison, we phase-folded $\approx4,\!000\,\mathrm{s}$ of simulation light curve data with an oscillation frequency of $\approx5.37\,\mathrm{mHz}$ (local frame) and took the average luminosity for each bin (\textsl{right axis}).  The luminosity is corrected by the assumed surface red shift, i.e., $L_\infty = L/(1+z)$, where $1+z=1/\sqrt{1-2GM/c^{2}R}\approx1.26$ for $1.4\,\mathrm{M}_{\odot}$ and $11.2\,\mathrm{km}$ radius neutron star, and then offset by a persistent luminosity $\approx 1.93\times10^{38}\,\mathrm{erg}\,\mathrm{s}^{-1}$ for $0.96\,\dot{M}_{\text{Edd}}$ local accretion rate with a $0.759$ hydrogen mass fraction.  For clarity, both $y$-axes offsets and scaling are adjusted arbitrarily for matching.  Both asymmetric shapes of the oscillations show a steeper rise than decay.  The fractional RMS amplitudes of the observed and the simulated are $\approx5.3\,\%$ and $0.65\,\%$, respectively.  This discrepancy is discussed in the following section.

\begin{table*}
	\centering
	\caption{Key figures and properties of $\mathrm{mHz}$ QPOs from different sources.}
	\label{summary}
    \begin{threeparttable}
	\begin{tabular}{lccccl} % four columns, alignment for each
		\hline
		 Source& \makecell{Accretion luminosity\\($10^{37}\,\mathrm{erg}\,\mathrm{s}^{-1}$)} & \makecell{QPOs frequency\\$(\mathrm{mHz})$} & \makecell{R.M.S\\$(\%)$} & \makecell{QPO Disappear\\after bursts\textsuperscript{a}} & References \\
		\hline
		4U~1636$-$536 & [0.6--$3.5]_{(2-150\,\mathrm{keV})}$ & 7--14.3 & [0.7--$1.9]_{(0.2-5\,\mathrm{keV})}$ & Y(D) & \makecell[l]{\cite{firstqpo} \\\cite{2008ApJ...673L..35A}}\\
		\makecell[l]{4U~1608$-$52 \\ Aql~X$-$1} & \makecell{[$\sim0.5$--$1.5]_{(3-20\,\mathrm{keV})}$ \\$-$} & \makecell{ $\sim7$--9 \\$\sim6$--7} & \makecell{[1.0--$2.1]_{(2-5\,\mathrm{keV})}$ \\$-$} & \makecell{Y \\$-$} & \cite{firstqpo}\\
		4U~1323$-$619 & $0.06_{(3-25\,\mathrm{keV})}\textsuperscript{b}$ & $8.1$ &$6.4_{(3-20\,\mathrm{keV})}$ & Y & \cite{2011ATel.3258....1S}\\
		IGR~J17480$-$2446 & [9--$12]_{(2-50\,\mathrm{keV})}$ & 2.8--4.2 & [1.3--$2.2]_{(2-60\,\mathrm{keV})}$ & ? & \cite{2012ApJ...748...82L}\\
		GS~1826$-$238 & $\sim2_{(0.6-9\,\mathrm{keV})}$ & $\sim8$ & [0.60--$0.95]_{(0.4-7.5\,\mathrm{keV})}$ &$-$&\cite{2018ApJ...865...63S}\\
		EXO~0748$-$676 & [$\sim0.25$--$0.75]_{(3-50\,\mathrm{keV})}$ & $\sim$5--13 & $\sim4_{(2-5\,\mathrm{keV})}$ & Y(D) &\cite{2019MNRAS.486L..74M}\\
		1RXS~J180408.9$-$342058 & $0.68_{(0.45-50\,\mathrm{keV})}\textsuperscript{c}$ & $\sim5$--8 &$5.3_{(0.7-10\,\mathrm{keV})}$ & Y & This work\\
		\hline
	\end{tabular}
	\begin{tablenotes}
      \item[a]$\text{D}=\,$Frequency of the oscillations systematically drops before being undetectable in some cases. Indeterminate for IGR~J17480-2446 as bursts and QPOs appear in separate observations.
      \item[b]Persistent flux obtained from \cite{minbarpaper}, assuming the source distance to be $4.2\,\mathrm{kpc}$ \citep{1323_distance}.
      \item[c]Persistent luminosity estimated by \cite{lum_for_1st_obs} with the assumed source distance $=5.8\,\mathrm{kpc}$.
    \end{tablenotes}
    \end{threeparttable}
\end{table*}
\section{Discussion and Conclusion}
% add non-break space (~) so that source names are not split across lines; good
% also the symbol in the names is a minus sign not hyphen, so use math mode
1RXS~J180408.9$-$342058 is the ninth source showing mHz QPOs from the NS LMXBs, along with 4U~1636$-$536 , 4U~1608$-$52 and Aql~X-1 \citep{firstqpo}, 4U~1323$-$619 \citep{2011ATel.3258....1S}, IGR~J17480$-$2446 \citep{2012ApJ...748...82L}, IGR~J00291+5934 \citep{2017MNRAS.466.3450F}, GS~1826$-$238 \citep{2018ApJ...865...63S} and EXO~0748$-$676 \citep{2019MNRAS.486L..74M}.  Except IGR~J00291+5934 \citep{2017MNRAS.466.3450F}, the presence of Type I bursts shows direct evidence to support the $\mathrm{mHz}$ QPOs being as a result of marginally stable burning. \cite{2017MNRAS.466.3450F} reported that the $\mathrm{mHz}$ signal found in IGR~J00291+5934 may be triggered by the so-called "heartbeat model" \citep{2011ApJ...742L..17A} which is associated with the movement of the inner accretion disk.  The correlated increases in the blackbody radius and the normalization of the Comptonization component, along with QPO flux in their QPO phase-resolved spectrum analysis support this attribution.  Moreover, the lack of Type I X-ray bursts during when the QPOs occur, and the very high RMS amplitude of QPOs ($\geq30\,\%$) also contra-indicate thermonuclear burning.  We summarized the key figures and properties of the marginally stable nuclear burning from the eight sources, excluding IGR~J00291+5934, in Table~\ref{summary}. It includes the accretion luminosity for the occurrence of mHz QPOs, the discovered frequency range and  fractional RMS amplitude of the oscillations, as well as whether the signal disappears after bursts.

In the first \textsl{XMM-Newton} observation of 1RXS~J180408.9$-$342058, when the mHz QPOs were present, the source was in a very hard state (photon index $\lesssim1.4$) at persistent luminosity $=6.8\times10^{36}\,\mathrm{erg}\,\mathrm{s}^{-1}$ \citep{lum_for_1st_obs}.  The source had transited to hard/island state, accompanied with $\approx2.5$ times increase in its persistent luminosity, nearly a month later, captured in the second \textsl{XMM-Newton} observation.  With this moderate change of accretion rate, the mHz oscillations became undetectable, along with the power-law noise in the low frequency range.  In contrast, in the previous detections of mHz QPOs, 4U~1636$-$536, IGR~J17480$-$2446, GS~1826$-$238, and EXO~0748$-$676 were reported to be in a relatively soft state, unlike 1RXS~J180408.9$-$342058.  Our result may imply that the occurrence of the marginally stable burning is not limited by the position of the source state. 

The disappearance of QPOs with the onset of Type I X-ray bursts was reported in the sources 4U 1608-52 \citep{firstqpo}, 4U~1323$-$619 \citep{2011ATel.3258....1S}, 4U~1636$-$536 \citep{2008ApJ...673L..35A,1636disappear}, and EXO~0748$-$676 \citep{2019MNRAS.486L..74M}.  For the latter two sources, it was also reported that the frequency of the oscillations has systematically dropped before it is undetectable. We found in 1RXS~J180408.9$-$342058 that the strength of the signals drops when approaching a burst, and then disappears afterward, though no obvious decreasing trend on the signal frequency prior to bursts was found.  

On the other hand, \cite{2012ApJ...748...82L} showed that mHz QPOs appear through a smooth evolution of burning regime in the slowest spinning neutron star, IGR~J17480$-$2446.  In that source, the burst rate increases along with the accretion rate until a transitional phase, where bursts disappear with the onset of periodic oscillations, and vice versa when the accretion rate declines leading to the drop of burst rate.  Furthermore, the accretion luminosity at which the $\mathrm{mHz}$ QPOs were observed is $\sim10^{38}\,\mathrm{erg}\,\mathrm{s}^{-1}$, which is closer to the theoretical value ($\sim\dot{M}_{\text{Edd}}$ by, e.g., \citealt{1998ASIC..515..419B}) compared to the reported range of luminosity $\sim 10^{36}$--$10^{37}\,\mathrm{erg}\,\mathrm{s}^{-1}$ in other sources.  Moreover, the presence of bursts and mHz QPOs are in separate observations with distinct luminosities, unlike  in the more rapidly-spinning sources, where the bursts and  oscillations occur within the same observations.    

Independent of the variations in mHz QPO properties, such as the accretion rate for the oscillations to exist as well as whether they disappear after bursts (refer to Table~\ref{summary}), remains some uncertainties as to whether the involved nuclear processes in different sources are the same.  In fact, the actual nuclear burning depends on various conditions such as fuel compositions, crustal heating, and neutron star rotation rate, etc. \cite{2018ApJ...865...63S} pointed out that fast rotating NS has a significant difference of surface gravity from the pole and equator.  For example, the $582\,\mathrm{Hz}$ rotating NS in the LMXB 4U~1636$-$536 has surface gravity at the pole $\approx11\,\%$ higher than at the equator.  
% "this" or "these" should always be followed by a noun!
This variation may lead to different burning regimes across the entire NS surface as they are sensitive to the surface gravity \citep{heger2007}.  Besides, \cite{2018ApJ...857L..24G} provided additional evidence in support of the influence of NS spin on the burning regime.  They found that the regime of maximum burst rate is at a lower accretion rate if the NS has a higher spin rate.  Once the burst rate reaches a maximum, it starts to decrease with further increases of accretion rate, supposedly because of the enhanced stabilization of nuclear burning \citep{van1988}.  Thus, the occurrence of mHz QPOs, triggered by marginally stable burning, shifts to a lower accretion rate for a NS with faster rotation.  Interestingly, IGR~J17480$-$2446, with only a $\mathrm{11\,\mathrm{Hz}}$ spin rate, is the slowest rotating NS in LMXBs, and also where the observational properties of the mHz QPOs match better with the 1D simulation results, in terms of the accretion rate for the occurrence of the marginally stable burning.  Therefore, we also speculate that the spin rate may play a predominant factor to the mentioned discrepancies between the observed mHz QPOs and theory.

% The energy dependent QPOs amplitude \ref{fig:energy_dependent} from the two dwell show increase toward lower energy. Similar result can be found on IGR J17480-2446 \citep{2012ApJ...748...82L} where they show the trend is relatively flat in $\sim 2 \text{ to } 6 \mathrm{keV}$ but increase up to $11 \mathrm{keV}$ based on \textit{RXTE} data. This is inconsistent to the finding by \citep{firstqpo} where they found stronger pulsation in $2\text{ to }4 \mathrm{keV}$ and no obvious increases in higher energy band. 

%, and also the wide range of persistent luminosity, from $10^{35}$--$10^{38}\,\mathrm{erg}\,\mathrm{s}^{-1}$, for the mHz QPOs to occur

\begin{figure}
	\includegraphics[width=\columnwidth]{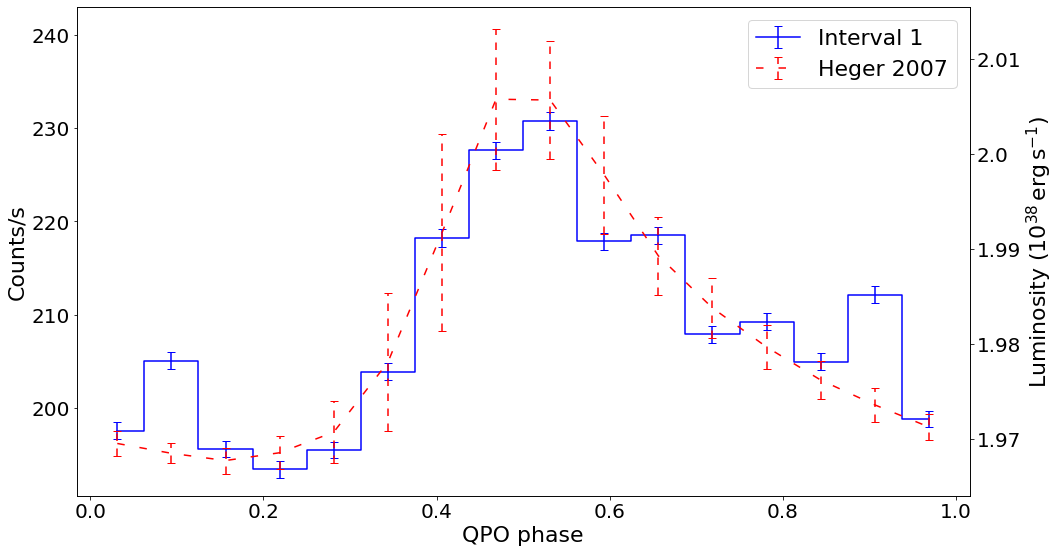}
    \caption{A comparison between the mHz QPOs profiles of 1RXS~J180408.9$-$342058 and simulated signals.  Events from the first $4,\!000\, \mathrm{s}$ interval in Figure~\ref{fig:lc} were folded with a frequency $6.57\,\mathrm{mHz}$ found in Figure~\ref{fig:confidence}.  A $\sim4,\!000\, \mathrm{s}$ light curve is extracted from simulation \citep{heger2007} and folded with a local oscillation frequency $\approx5.37\, \mathrm{mHz}$. The average luminosity from the folded light curve is corrected by surface red shift, and offset by a persistent value (plotted referenced to the right axis). Both $y$-axes offsets and scaling are adjusted arbitrarily for matching. Results from the observed and simulation similarly show a shape with a steep rise and shallower decline, while the oscillation amplitudes are notably different.}
    \label{fig:overplot}
\end{figure}

In Figure~\ref{fig:overplot} we compare the QPO shapes between the observed ($4,\!000\,\mathrm{s}$ interval from the first segment) and the simulation result from \cite{heger2007}.  Both results show asymmetric oscillation profiles with a steeper rise than decay.  We speculate that the shape of the oscillations is mainly dependent on the burning fuel compositions.  On the other hand, their fractional RMS amplitudes are distinctly different, $\approx5.3\,\%$ from the observed whereas $\approx0.65\,\%$ from the simulation.  For gravitational energy released from accretion $\approx200\,\mathrm{MeV}$ comparing to the energy released from thermonuclear fusion $\approx 5\,\mathrm{MeV}$ per nucleon, we would expect a maximum theoretical oscillation amplitude of $5/200=2.5\,\%$; in the simulations \citep{heger2007}, however, the maximum oscillation amplitude is only half as big but non-sinusoidal, implying an upper limit for a nuclear-powered oscillation amplitude of only $\sim1.5\,\%$.  This reduction is due to the partial energy contribution to the steady state burning.  Hence thermonuclear burning by itself seemingly cannot be responsible for such a large oscillation amplitude.  More comprehensive simulation studies, taking into account the full geometrical and general relativistic effects, are required to confirm this deficiency.
% We found that there is another observation shows 'clocked burst' but no any significant periodic signals are detected. 

\section*{Acknowledgements}
Based on observations obtained with \textsl{XMM-Newton}, an ESA science mission with instruments and contributions directly funded by ESA Member States and NASA.  This work was supported in part by the National Science Foundation under Grant No.\ PHY-1430152 (JINA Center for the Evolution of the Elements).
Parts of this research were conducted by the   Australian Research Council Centre of Excellence for Gravitational Wave Discovery (OzGrav), through project number CE170100004.  
AH has been supported, in part by the Australian Research Council Centre of Excellence for All Sky Astrophysics in 3 Dimensions (ASTRO 3D), through project number CE170100013.  Y.C. and H. -E. H.  especially acknowledges the support from Ministry of Science and Technology of Taiwan through grant MOST 109-2112-M-008-004-.
%AH has been supported by a grant from Science and Technology Commission of Shanghai Municipality (Grants No.16DZ2260200) and National Natural Science Foundation of China (Grants No.11655002).  
\section*{Data availability}
The \textsl{XMM-Newton} data underlying this article are available in the High Energy Astrophysics Science Archive Research Center (HEASARC), at (\url{https://heasarc.gsfc.nasa.gov}).

%%%%%%%%%%%%%%%%%%%%%%%%%%%%%%%%%%%%%%%%%%%%%%%%%%



\bibliographystyle{mnras}
\bibliography{ms} % if your bibtex file is called example.bib

%\begin{thebibliography}{99}
%\end{thebibliography}

%%%%%%%%%%%%%%%%%%%%%%%%%%%%%%%%%%%%%%%%%%%%%%%%%%

%%%%%%%%%%%%%%%%% APPENDICES %%%%%%%%%%%%%%%%%%%%%

%%%%%%%%%%%%%%%%%%%%%%%%%%%%%%%%%%%%%%%%%%%%%%%%%%

% Don't change these lines
\bsp	% typesetting comment
\label{lastpage}
\end{document}